\documentstyle[aps,prb,epsfig,float,twocolumn]{revtex}
\begin{document}
\twocolumn[\hsize\textwidth\columnwidth\hsize\csname
@twocolumnfalse\endcsname

\title{Theoretical study of interacting hole gas in p-doped bulk III-V
    semiconductors}

\author{John Schliemann}

\address{Institute for Theoretical Physics, University of Regensburg, 
D-93040 Regensburg, Germany}

\date{\today}

\maketitle

\begin{abstract}
We study the homogeneous interacting hole gas in 
$p$-doped bulk III-V semiconductors.
The structure of the valence band is modelled by Luttinger's Hamiltonian
in the spherical approximation, giving rise to heavy and light hole dispersion
branches, and the Coulomb repulsion is taken into account via a self-consistent
Hartree-Fock treatment. As a nontrivial feature of the model, the 
self-consistent solutions of the Hartree-Fock equations 
can be found in an almost
purely analytical fashion, which is not the case for other types of effective
spin-orbit coupling terms. In particular, the Coulomb interaction 
renormalizes the Fermi wave numbers for heavy and light holes. As a
consequence, the ground state energy found in the self-consistent Hartree-Fock
approach and the result from lowest-order perturbation theory do not agree.
We discuss the consequences of our observations for ferromagnetic 
semiconductors, and for the possible observation of the spin-Hall effect in
bulk $p$-doped semiconductors. Finally, we also investigate elementary
properties of the dielectric function in such systems.
\end{abstract}
\vskip2pc]

\section{Introduction}

Over the last years, effects of spin-orbit coupling in semiconductors have 
moved into the very focus of both experimental and theoretical
solid-state research, mainly within the large and still rapidly growing field
of spintronics \cite{overview}. In the p-type valence band of III-V zinc-blende
semiconductors spin-orbit interaction is particularly strong. Important 
examples of $p$-doped semiconductor systems with itinerant charge carriers in
the valence band include ferromagnetic semiconductors with 
${\rm Ga}_{1-x}{\rm Mn}_x{\rm As}$ being the most intensively studied material,
for an overview see Refs.~\cite{Timm03,MacDonald05,Jungwirth06}. In 
${\rm Ga}_{1-x}{\rm Mn}_x{\rm As}$ and related systems, the substitutional
Mn dopants form local
moments with spin $S=5/2$ from its five d-electrons, while they also act as
acceptors providing holes in the valence band interacting with the local 
moments. This interaction between charge carriers and local spin moments
then leads, at low enough temperatures, to 
ferromagnetic order, giving rise to the notion of {\em carrier-induced
ferromagnetism}. So far, Curie temperatures as high as 
$T_{c}\approx 160\dots 170{\rm K}$ have 
been observed \cite{Edmonds04,Jungwirth05}.
There is a vast literature on the 
theoretical description of ferromagnetic semiconductors
taking into account realistic
band structure parameterizations for the valence band, for an early key 
publication see Ref.~\cite{Dietl00}.
However, what is most often neglected in the treatment of models for 
ferromagnetic semiconductors is the Coulomb interaction
{\em among} the holes. A (semi-)phenomenological way to account for
Coulomb repulsion is to introduce appropriate Fermi liquid parameters
\cite{Dietl01}.
Exceptions to these heuristic approaches
include numerical work based on dynamical
mean field theory \cite{Craco03}, and
a numerical Hartree-Fock study of a disordered two-band model, neglecting
spin-orbit coupling \cite{Yang03}. 
For further dynamical-mean-field studies of models for ferromagnetic 
semiconductors not incorporating Coulomb repulsion see 
Refs.~\cite{Chattopadhyay01,Aryanpour05,Popescu06}.

Moreover, $p$-doped semiconductors have also attracted interest 
with respect to the recently predicted intrinsic spin Hall effect 
\cite{Murakami03,Sinova03,Schliemann05}; for a recent overview see also
Ref.~\cite{Schliemann06}. In fact, 
the pioneering paper by Murakami, Nagaosa, and
Zhang studies a $p$-doped bulk III-V semiconductor taking into account 
heavy and light hole bands around the $\Gamma$-point \cite{Murakami03}.
However, the Coulomb repulsion between holes was also neglected here. 

In summary, in the light of the above challenges and activities, it is 
certainly desirable to
develop a deeper and possibly least partially analytical understanding
of the effects of Coulomb interaction in $p$-doped semiconductors taking 
into account spin-orbit coupling. In the 
present work we study interacting holes
in the valence band of a III-V semiconductor. The band structure is modelled by
Luttinger's Hamiltonian in the spherical approximation leading to
heavy and light hole dispersion branches \cite{Luttinger56}.
The Coulomb repulsion between holes is treated via Hartree-Fock theory. 
As a nontrivial feature of the model, the 
self-conistent solutions of the Hartree-Fock equations can be found in an 
almost
purely analytical fashion, which is not the case for other types of effective
spin-orbit coupling terms. In particular, the Coulomb interaction 
renormalizes the Fermi wave numbers for heavy and light holes. As a
consequence, the ground state energy found in the self-consistent Hartree-Fock
approach and the result from lowest-order perturbation theory do not agree.
In other words, the self-consistent Hartree-Fock treatment contains
contributions beyond lowest-order perturbation theory, which is a
result of the nontrivial band structure.
We discuss the consequences of our observations for ferromagnetic 
semiconductors, and for the possible observation of the spin-Hall effect in
bulk $p$-doped semiconductors. Moreover, we also investigate elementary
properties of the dielectric function in such systems.

This paper is organized as follows. In section \ref{nonint} we introduce the
single-particle Hamiltonian and basic properties of the non-interacting system.
In particular, the structure of the single-particle eigenstates will be of 
importance for the Hartree-Fock study in section \ref{HF}. The self-consistent
solution of the Hartree-Fock equations for Coulomb repulsion is presented
in section \ref{coulomb}. In section \ref{comparison} we compare our findings
for the three-dimensional hole gas with the situation in other generic
semiconductor structures where spin-orbit coupling plays an important role.
We then return to the three-dimensional hole gas and investigate its
ground state energy and pair correlations functions in Hartree-Fock 
theory. In section \ref{RPA} we also discuss elementary properties of
the dielectric function within random phase approximation. We close with a 
discussion and outlook in section \ref{discussion}.

\section{The non-interacting hole gas}
\label{nonint}

A good approximative description of heavy and light hole states around the
$\Gamma$-point in III-V zinc-blende semiconductors is given by Luttinger's
 Hamiltonian \cite{Luttinger56},
\begin{equation}
{\cal H}=\frac{1}{2m_{0}}\left(\left(\gamma_{1}+\frac{5}{2}\gamma_{2}\right)
\vec p^{2}-2\gamma_{2}\left(\vec p\cdot\vec S\right)^{2}\right)\,.
\label{Luttinger}
\end{equation}
Here $m_{0}$ is 
the bare electron mass, $\vec p$ is the hole lattice momentum, 
and $\vec S$ are spin-$3/2$-operators, resulting from
adding the $l=1$ orbital angular momentum to the $s=1/2$ electron spin.
The dimensionless Luttinger parameters $\gamma_{1}$ and $\gamma_{2}$ 
describe the valence
band of the specific material with effects of spin-orbit coupling being
included in $\gamma_{2}$. 
The above Hamiltonian is rotationally invariant and commutes with the 
helicity operator $\lambda=(\vec k\cdot\vec S)/k$, where $\vec k=\vec p/ \hbar$ 
is the hole wave vector. 

Thus, the eigenstates of (\ref{Luttinger})
can be chosen to be eigenstates of the helicity operator, which grossly
facilitates analytical calculations.
The heavy holes correspond to
$\lambda=\pm 3/2$, while the light holes have $\lambda=\pm 1/2$. For the dispersions
of $\varepsilon_{h/l}(\vec k)$ of heavy and light hole states, respectively,
one finds
\begin{equation}
\varepsilon_{h/l}(\vec k)=\frac{\hbar^{2}k^{2}}{2m_{h/l}}
\end{equation}
where the masses $m_{h/l}$ of heavy and light holes are given by
\begin{equation}
m_{h/l}=\frac{m_{0}}{\gamma_{1}\mp 2\gamma_{2}}\,.
\end{equation}
Well established values for
the Luttinger parameters, among other band structure parameters, can be found 
in the literature \cite{Vurgaftman01}. For example, for GaAs one has
$\gamma_{1}\approx 7.0$ and $\gamma_{2}\approx 2.5$ giving $m_{h}\approx 0.5m_{0}$ and $m_{l}\approx 0.08m_{0}$.

The corresponding eigenstates of the Hamiltonian (\ref{Luttinger}) are given
by
\begin{equation}
\langle\vec r|\vec k,\lambda\rangle=\frac{e^{i\vec k\vec r}}{\sqrt{V}}|\chi_{\lambda}(\vec k)\rangle\,,
\label{eigenstate}
\end{equation}
where $V$ is the volume of the system. Using the conventional basis of 
eigenstates of $S^{z}$ and introducing the usual parameterization
$\vec k=k(\cos\varphi\sin\vartheta,\sin\varphi\sin\vartheta,\cos\vartheta)$ in terms of polar coordinates,
the eigenspinors $|\chi_{\lambda}(\vec k)\rangle$ of the helicity operator
$\lambda=(\vec k\cdot\vec S)/k$ read explicitly  
\begin{eqnarray}
|\chi_{\frac{3}{2}}(\vec k)\rangle & = & \left(
\begin{array}{c}
\cos^{3}\frac{\vartheta}{2}e^{-\frac{3i}{2}\varphi}\\
\sqrt{3}\cos^{2}\frac{\vartheta}{2}\sin\frac{\vartheta}{2}e^{-\frac{i}{2}\varphi}\\
\sqrt{3}\cos\frac{\vartheta}{2}\sin^{2}\frac{\vartheta}{2}e^{+\frac{i}{2}\varphi}\\
\sin^{3}\frac{\vartheta}{2}e^{+\frac{3i}{2}\varphi}
\end{array}
\right)\label{spinor1}\\
|\chi_{\frac{1}{2}}(\vec k)\rangle & = & \left(
\begin{array}{c}
-\sqrt{3}\cos^{2}\frac{\vartheta}{2}\sin\frac{\vartheta}{2}e^{-\frac{3i}{2}\varphi}\\
\cos\frac{\vartheta}{2}\left(\cos^{2}\frac{\vartheta}{2}
-2\sin^{2}\frac{\vartheta}{2}\right)e^{-\frac{i}{2}\varphi}\\
\sin\frac{\vartheta}{2}\left(2\cos^{2}\frac{\vartheta}{2}
-\sin^{2}\frac{\vartheta}{2}\right)e^{+\frac{i}{2}\varphi}\\
\sqrt{3}\cos\frac{\vartheta}{2}\sin^{2}\frac{\vartheta}{2}e^{+\frac{3i}{2}\varphi}
\end{array}
\right)
\label{spinor2}
\end{eqnarray}
and the remaining eigenspinors $|\chi_{-3/2}(\vec k)\rangle$, $|\chi_{-1/2}(\vec k)\rangle$ can be
obtained from the above ones by shifting $\vartheta\mapsto\pi-\vartheta$, $\varphi\mapsto\varphi+\pi$, corresponding
to a spatial inversion $\vec k\mapsto -\vec k$.
Note that $|\chi_{\pm\frac{3}{2}}(\vec k)\rangle$ is just a usual 
spin-coherent state of spin length $S=3/2$ polarized along the
direction $\pm\vec k/k$.
In what follows, we will also need the mutual overlaps squared
between spinors which are given by
\begin{eqnarray}
|\langle\chi_{\frac{3}{2}}(\vec k_{1})|\chi_{\frac{3}{2}}(\vec k_{2})\rangle|^{2} & = &
\left(\frac{1}{2}\left(1+\frac{\vec k_{1}\vec k_{2}}{k_{1}k_{2}}\right)\right)^{3}
\label{overlap1}\\
|\langle\chi_{\frac{1}{2}}(\vec k_{1})|\chi_{\frac{1}{2}}(\vec k_{2})\rangle|^{2} & = &
\frac{1}{8}\left(1+\frac{\vec k_{1}\vec k_{2}}{k_{1}k_{2}}\right)
\left(3\frac{\vec k_{1}\vec k_{2}}{k_{1}k_{2}}-1\right)^{2}
\label{overlap2}\\
|\langle\chi_{\frac{3}{2}}(\vec k_{1})|\chi_{\frac{1}{2}}(\vec k_{2})\rangle|^{2} & = &
\frac{3}{8}\left(1+\frac{\vec k_{1}\vec k_{2}}{k_{1}k_{2}}\right)^{2}
\left(1-\frac{\vec k_{1}\vec k_{2}}{k_{1}k_{2}}\right)
\label{overlap3}
\end{eqnarray}
These expressions can be derived easily from Eqs.~(\ref{spinor1}),
(\ref{spinor2}) by putting one of the wave vectors along the $z$-direction
and writing the resulting  overlap squared in an explicitly rotationally
invariant fashion as above. 

Let us now consider a non-interacting hole gas in an infinite system. 
Then the ground state is
characterized by the Fermi wave numbers
\begin{equation}
k_{h/l}=\sqrt{\frac{2m_{h/l}}{\hbar^{2}}\varepsilon_{f}}
\end{equation}
where $\varepsilon_f$ is the Fermi energy. These wave numbers are related to the density
$n=N/V$, $N$ being the number of holes, via
\begin{equation}
n=\frac{1}{3\pi^{2}}\left(k_{h}^{3}+k_{l}^{3}\right)\,.
\end{equation}
The kinetic energy per particle is straightforwardly obtained as
\begin{equation}
\frac{E_{kin}}{N}=\frac{1}{5\pi^{2}n}
\left(\frac{\hbar^{2}}{2m_{h}}k_{h}^{5}+\frac{\hbar^{2}}{2m_{l}}k_{l}^{5}\right)\,.
\end{equation}
The above expression suggests to introduce an averaged mass $\tilde m$ by
defining
\begin{equation}
\tilde m^{\frac{3}{2}}=\frac{1}{2}\left(m_{h}^{\frac{3}{2}}+m_{l}^{\frac{3}{2}}\right)
\end{equation}
along with an averaged Fermi wave number
\begin{equation}
\tilde k=\sqrt{\frac{\tilde m}{m_{h/l}}}k_{h/l}
\end{equation}
fulfilling
\begin{equation}
n=\frac{2}{3\pi^{2}}\tilde k^{3}\
\end{equation}
and 
\begin{equation}
\tilde k=\sqrt{\frac{2\tilde m}{\hbar^{2}}\varepsilon_{f}}\,.
\end{equation}
The kinetic energy per particle can then be rewritten as
\begin{equation}
\frac{E_{kin}}{N}=\frac{3}{10}\frac{\hbar^2}{\tilde m}\tilde k^{2}\,,
\end{equation}
which exactly resembles the familiar expression for the usual spin-$1/2$
electron gas \cite{Mahan00,Rossler04}. 
In circumstances of Coulomb interaction between the holes,
the above finding suggests to introduce a density parameter
$r_{s}$ and a Bohr radius $\tilde a_{B}$ by defining
\begin{equation}
\frac{1}{n}=\frac{4\pi}{3}\left(r_{s}\tilde a_{B}\right)^{3}
\label{denpar}
\end{equation}
and
\begin{equation} 
\tilde a_{B}=\frac{\hbar^{2}}{\tilde me^{2}}\varepsilon_{r}\,,
\label{bohr}
\end{equation}
where $e$ is the electron charge, and we have introduced a static
dielectric constant $\varepsilon_{r}$ to account for screening from electrons
in remote bands. Then the kinetic energy per particle can be rewritten as
\begin{equation}
\frac{E_{kin}}{N}=\frac{3}{5}\left(\frac{9\pi}{8}\right)^{2/3}\frac{1}{r_{s}^{2}}
\frac{e^2}{2\tilde a_{B}\varepsilon_{r}}\,,
\end{equation}
where $e^2/2\tilde a_{B}\varepsilon_{r}$ is the Rydberg energy unit. Up to a slight difference 
in the prefactor, the above expression is
again completely analogous to the result for the usual electron gas
\cite{Mahan00,Rossler04}.
However, as we shall see below, the exchange contribution from Coulomb 
interaction cannot be casted in a form immediately analogous
to the spin-degenerate electron gas.

\section{The interacting hole gas in self-consistent
Hartree-Fock approximation}
\label{HF}

We now consider an infinite system with a repulsive interaction between
the holes which is naturally assumed to be translationally and rotationally
invariant. Later on it will be specified to be the Coulomb repulsion. Moreover,
we assume a homogeneous neutralizing background ensuring charge
neutrality and cancelling all direct (or Hartree) contributions from
Hartree-Fock expressions.

The eigenstates (\ref{eigenstate}) of the single-particle
Hamiltonian (\ref{Luttinger}) solve the Hartree-Fock equations
\begin{eqnarray}
 & & \varepsilon^{HF}_{\lambda}(\vec k)\langle\vec r|\vec k,\lambda\rangle=\frac{\hbar^{2}k^{2}}{2m_{\lambda}}
\langle\vec r|\vec k,\lambda\rangle\nonumber\\
& & \qquad-\frac{e^{i\vec k\vec r}}{\sqrt{V}}\frac{1}{(2\pi)^{3}}\sum_{\lambda'}\int_{k\leq q_{\lambda'}} d^{3}k'
\langle\chi_{\lambda'}(\vec k')|\chi_{\lambda}(\vec k)\rangle\nonumber\\
& & \qquad\qquad\qquad \cdot V(|\vec k-\vec k'| )|\chi_{\lambda'}(\vec k')\rangle\,,
\label{HFequation}
\end{eqnarray}
where $V( |\vec k| )$ is the Fourier transform of the interaction potential,
and $m_{\lambda}$ stands for $m_{h}$ ($m_{l}$) if $\lambda=\pm 3/2$ ($\lambda=\pm 1/2$), a notation scheme
which we will also use in the following. The Hartree-Fock eigenenergies
are given by
\begin{eqnarray}
 & & \varepsilon^{HF}_{h/l}(\vec k;q_{h},q_{l})=\frac{\hbar^{2}k^{2}}{2m_{h/l}}\nonumber\\
 & & -\frac{1}{(2\pi)^{3}}\int_{k'\leq q_{h/l}} d^{3}k'\frac{1}{4}
\left(1+3\left(\frac{\vec k\vec k'}{kk'}\right)^{2}\right)
V(|\vec k-\vec k'| )\nonumber\\
 & & -\frac{1}{(2\pi)^{3}}\int_{k'\leq q_{l/h}} d^{3}k'\frac{3}{4}
\left(1-\left(\frac{\vec k\vec k'}{kk'}\right)^{2}\right)
V(|\vec k-\vec k'| )\,.
\label{HFenergy1}
\end{eqnarray}
To see that the eigenstates (\ref{eigenstate}) solve the above
Hartree-Fock equations, one can, again without loss of generality, 
take the wave vector $\vec k$ in Eq.~(\ref{HFequation}) to point along the
$z$-direction. Using the explicit parameterizations (\ref{spinor1}),
(\ref{spinor2}) of the eigenspinors in terms of polar coordinates, one
easily sees that the integration over the azimuthal angle $\varphi'$ ensures
that the integral in Eq.~(\ref{HFequation}) is indeed proportional
to $|\chi_{\lambda}(\vec k)\rangle$. This result holds for any interaction potential
since $|\vec k-\vec k'|$ is independent of $\varphi'$.
The eigenvalues (\ref{HFenergy1}) are then derived by
performing the summation over $\lambda'$ in Eq.~(\ref{HFequation}) and using
Eqs.~(\ref{overlap1})-(\ref{overlap3}). The first integral
in Eq.~(\ref{HFenergy1}) stems from the contributions with
$|\lambda|=|\lambda'|$ where as the second integral results from the cases
$|\lambda|\neq|\lambda'|$. As explicitly shown in the
notation of Eq.~(\ref{HFenergy1}), these eigenvalues are functions of
the integration boundaries $q_{h}$, $q_{l}$ arising in Eq.~(\ref{HFequation}).
In the presence of interactions, these quantities will in general not coincide
with the Fermi wave numbers $k_{h}$, $k_{l}$ of the non-interacting system,
as we shall see below.

\subsection{Self-consistent Hartree-Fock solution for Coulomb interaction:
renormalization of Fermi wave numbers}
\label{coulomb}

Let us now specify the interaction to be the Coulomb repulsion, i.e.
\begin{equation}
V(k)=\frac{e^{2}}{\varepsilon_{r}}\frac{4\pi}{k^{2}}\,.
\end{equation}
The Hartree-Fock eigenenergies read explcitly
\begin{eqnarray}
 & & \varepsilon^{HF}_{h/l}(\vec k;q_{h},q_{l})=\frac{\hbar^{2}k^{2}}{2m_{h/l}}\nonumber\\
 & & -\frac{e^{2}}{\varepsilon_{r}}\frac{q_{h/l}}{4\pi}\Biggl[\frac{11}{8}
-\frac{3}{8}\frac{q_{h/l}^{2}}{k^{2}}
+\frac{3}{4}\frac{k}{q_{h/l}}h\left(\frac{q_{h/l}}{k}\right)\nonumber\\
 & & \quad +\frac{k}{q_{h/l}}\left(\frac{5}{4}\frac{q_{h/l}^{2}-k^{2}}{k^{2}}
+\frac{3}{16}\frac{q_{h/l}^{4}-k^{4}}{k^{4}}\right)\log\left|\frac{q_{h/l}+k}{q_{h/l}-k}\right|
\Biggr]\nonumber\\
 & & -\frac{e^{2}}{\varepsilon_{r}}\frac{q_{l/h}}{4\pi}\Biggl[\frac{21}{8}
+\frac{3}{8}\frac{q_{l/h}^{2}}{k^{2}}
-\frac{3}{4}\frac{k}{q_{l/h}}h\left(\frac{q_{l/h}}{k}\right)\nonumber\\
 & & \quad +\frac{k}{q_{l/h}}\left(\frac{3}{4}\frac{q_{l/h}^{2}-k^{2}}{k^{2}}
-\frac{3}{16}\frac{q_{l/h}^{4}-k^{4}}{k^{4}}\right)\log\left|\frac{q_{l/h}+k}{q_{l/h}-k}\right|
\Biggr]\,,\nonumber\\
\label{HFenergy2}
\end{eqnarray}
where the function $h(x)$ is defined by
\begin{equation}
h(x)=\left\{
\begin{array}{ll}
2\sum_{n=0}^{\infty}\frac{x^{2n+1}}{(2n+1)^{2}} & x\leq 1 \\
\frac{\pi^{2}}{2}-2\sum_{n=0}^{\infty}\frac{\left(\frac{1}{x}\right)^{2n+1}}{(2n+1)^{2}} & x\geq 1
\end{array}
\right.\,.
\end{equation}
Note that $h(1)$ is simply related to Riemann's $\zeta$-function,
$h(1)=(3/2)\zeta(2)=\pi^{2}/4$.

\begin{figure}
\centerline{\includegraphics[width=7cm]{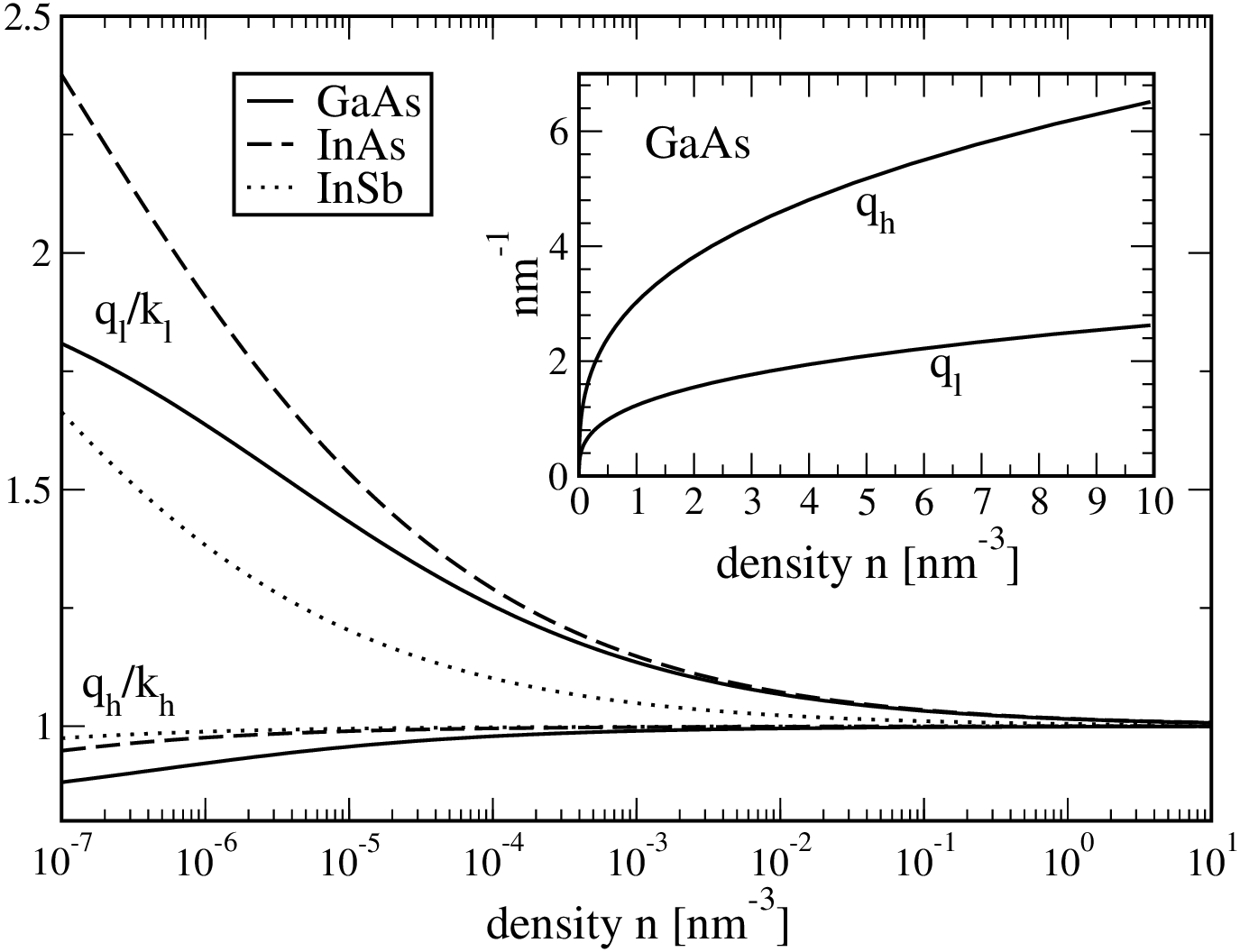}} 
\caption{The ratios of renormalized and unrenormalized 
Fermi wave numbers, $q_{h/l}$ and $k_{h/l}$, respectively, as a function of hole 
density for different III-V semiconductors. 
The renormalization of Fermi wave numbers affects primarily the
light hole wave number at low densities. 
The inset shows $q_{h}$ and $q_{l}$ as a function of density for GaAs.
\label{fig1}}
\end{figure}

The two dispersions branches (\ref{HFenergy2}) coincide at
zero wave vector,
\begin{equation}
\varepsilon^{HF}_{h/l}(0;q_{h},q_{l})=-\frac{e^{2}}{\varepsilon_{r}\pi}\left(q_{h}+q_{l}\right)
\end{equation}
for any values of $q_{h}$, $q_{l}$. However, when evaluated for the
Fermi wave numbers $k_{h}$, $k_{l}$, they differ at the corresponding wave numbers,
\begin{equation}
\varepsilon^{HF}_{h}(k_{h};k_{h},k_{l})\neq \varepsilon^{HF}_{l}(k_{l};k_{h},k_{l})\,.
\end{equation}
Of course the Fermi energies for heavy and light holes have to be the same 
since otherwise a redistribution of occupation numbers would take place.
Therefore, in order to obtain a truly self-consistent solution
to the Hartree-Fock equations, the Fermi wave numbers
$q_{h}$, $q_{l}$ have to be adjusted such that
\begin{equation}
\varepsilon^{HF}_{h}(q_{h};q_{h},q_{l})=\varepsilon^{HF}_{l}(q_{l};q_{h},q_{l})
\end{equation}
under the constraint of a fixed density,
\begin{equation}
n=\frac{1}{3\pi^{2}}\left(q_{h}^{3}+q_{l}^{3}\right)\,.
\label{density}
\end{equation}
Thus, in fact just a single parameter, say $q_{h}$, has to be determined
numerically, which is technically a very simple task. Fig.~\ref{fig1} shows
The ratios $q_{h/l}/k_{h/l}$ of renormalized to unrenormalized 
Fermi wave numbers as a function of hole 
density for the III-V semiconductors GaAs, InAs, and InSb. The relevant
parameters for these materials are summarized in table~\ref{table1}. 
As seen from the figure, for realistic parameters one always has
$q_{h}<k_{h}$ and $q_{l}>k_{l}$, i.e. due to Coulomb interaction heavy hole states
get depopulated in favor of light hole states. Moreover, the
renormalization of Fermi wavenumbers affects primarily the
light hole wave number at low densities. The inset of Fig.~\ref{fig1} 
shows $q_{h}$ and $q_{l}$ as a function of density for GaAs.

\begin{figure}
\centerline{\includegraphics[width=7cm]{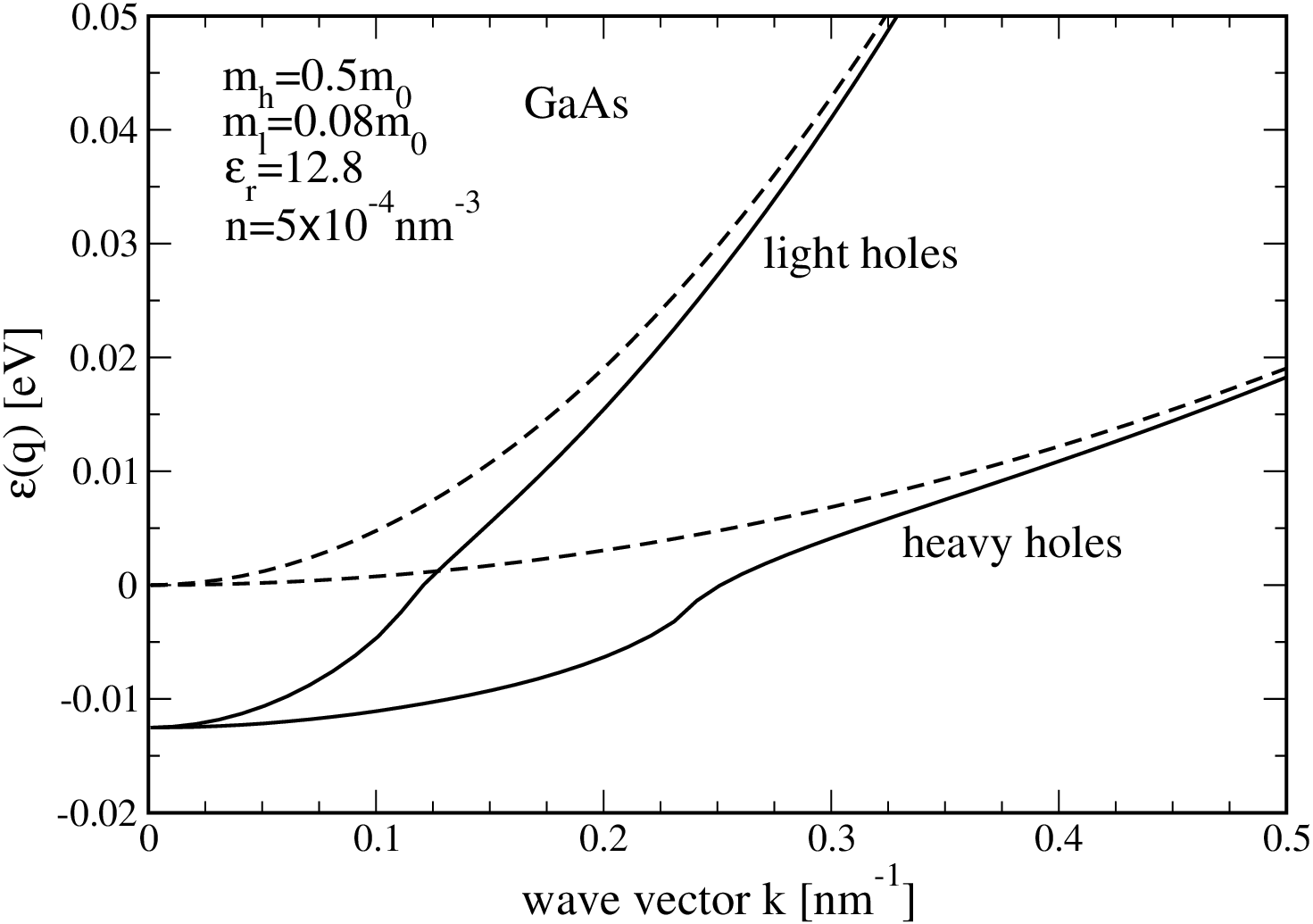}} 
\caption{The Hartree-Fock dispersions $\varepsilon^{HF}_{h/l}(k;q_{h},q_{l})$ for GaAs at a hole
density of $n=5\cdot 10^{-4}{\rm nm}^{-3}$. The solid lines show the dispersion
including Coulomb exchange for renormalized Fermi wave number $q_{h/l}$, while
the dashed lines represent the dispersions of the free hole gas in the absence
of interactions. The singularities in $\varepsilon^{HF}_{h/l}(k;q_{h},q_{l})$ at $k=q_{h/l}$ are clearly
pronounced while the singularities at at $k=q_{l/h}$ are weaker and hardly 
visible in the plot. 
\label{fig2}}
\end{figure}

In Fig.~\ref{fig2} we have plotted  
the Hartree-Fock dispersions $\varepsilon^{HF}_{h/l}(k;q_{h},q_{l})$ for GaAs at a hole
density of $n=5\cdot 10^{-4}{\rm nm}^{-3}$. The solid lines show the dispersion
including Coulomb exchange for renormalized Fermi wave number $q_{h/l}$, while
the dashed lines represent the dispersions of the free hole gas in the absence
of interactions. As seen from Eqs.~(\ref{HFenergy2}) the first derivative
of the dispersions $\varepsilon^{HF}_{h/l}(k;q_{h},q_{l})$ with respect to $k$ diverges both at
$k=q_{h}$ and $k=q_{l}$ giving rise to a vertical tangent at these
points. In Fig.~\ref{fig2} these singularities are clearly pronounced for
$\varepsilon^{HF}_{h/l}(k;q_{h},q_{l})$ at $k=q_{h/l}$ while they are weaker and hardly visible
in the plot at $k=q_{l/h}$. The fact that weak singularities occur 
in the derivative $\varepsilon^{HF}_{h/l}(k;q_{h},q_{l})$ also at $k=q_{l/h}$ is due to the mixing 
of heavy and light holes, i.e. the mutual overlap bet\-ween 
heavy and light
hole states at different wave vectors. Such an effect would be absent if one
just had two spin-1/2-species of different mass, say electrons and muons,
living in strictly different Hilbert spaces.

\subsection{Comparison with the two-dimensional electron gas and other systems}
\label{comparison}

As seen in Eq.~(\ref{HFequation}), the eigenstates (\ref{eigenstate}) 
of the non-interacting system provide solutions to the Hartree-Fock 
equations for a general pair interaction. This observation is 
familiar from the usual spin-1/2 electron gas without spin-orbit coupling
\cite{Mahan00,Rossler04}. If spin-orbit interaction is present, however,
such a simple structure cannot be taken for granted, and the solutions to the
Hartree-Fock equations can in general become more complicated. As an
example, consider a two-dimensional electron gas in a quantum well
being subject to Rashba spin-orbit coupling \cite{Rashba60},
\begin{equation}
{\cal H}=\frac{\vec p^{2}}{2m}+\frac{\alpha}{\hbar}\left(p_{x}\sigma^{y}-p_{y}\sigma^{x}\right)\,,
\label{rashba}
\end{equation}
where $m$ is an effective band mass, and $\alpha$ is the Rashba parameter
being tunable by an electric gate across the quantum well. 
$\vec\sigma$ are the usual Pauli matrices describing the electron spin.
We note that many-body effects in this type of system have recently 
attracted considerable interest
\cite{Chen99,Dimitrova04,Shekter05,Saraga05,Punnoose06,Pletyukhov05,Farid06}.
The above Hamiltonian has
two energy branches,
\begin{equation}
\varepsilon_{\pm}(\vec k)=\frac{\hbar^{2}k^{2}}{2m}\pm\alpha k
\end{equation}
with eigenstates
\begin{equation}
\langle\vec r|\vec k,\pm\rangle
=\frac{e^{i\vec k\vec r}}{\sqrt{A}}\frac{1}{\sqrt{2}}
\left(\begin{array}{c}
 1 \\ \pm (-k_{y}+ik_{x})/k
\end{array}\right)\,,
\label{rashbaeigenstate}
\end{equation}
where $A$ is the area of the system. Now consider an arbitrary pair 
interaction. Since the total Hamiltonian is still translationally
invariant, the solutions to the Hartree-Fock equations can always be chosen
to have a form similar as above, i.e. a plane wave factor times a 
two-component spinor. However, it is easy to see that the eigenstates
(\ref{rashbaeigenstate}) do not provide solutions to the Hartree-Fock equations
for a general interaction potential. This observation is due to the fact
that the angular integration in this two-dimensional case is different
from the three-dimensional situation of Eq.~(\ref{HFequation}). Only for
a pure contact interaction (having a constant Fourier transform), the
eigenstates (\ref{rashbaeigenstate}) solve the Hartree-Fock equations.

The same conclusions apply to bulk valence-band electrons being subject
to the three-dimensional Dresselhaus spin-orbit coupling term
\cite{Dresselhaus55}, and to heavy holes in asymmetric quantum wells
\cite{Winkler00,Gerchikov92}. In both cases, the effective spin-orbit 
interaction is trilinear in the particle momentum.

The situation of the two-dimensional electron gas 
becomes even more complicated if also Dresselhaus spin orbit
coupling is considered which reads in its
two-dimensional approximation \cite{Dyakonov86}
\begin{equation}
{\cal H}_{D}=\frac{\beta}{\hbar}\left(p_{y}\sigma^{y}-p_{x}\sigma^{x}\right)
\label{dressel}
\end{equation}
with a coupling parameter $\beta$. However, in the case when the Rashba
parameter is of equal magnitude as the Dresselhaus parameter,
$\alpha=\pm\beta$, the corresponding eigenstates of the non-interacting system
solve the Hartree-Fock equations for an arbitrary interaction potential.
This is due to the additional conserved quantity arising at this point
\cite{Schliemann03a} which cancels the effects of spin-orbit coupling
in many respects \cite{Schliemann03a,Schliemann03b}.

\subsection{Total ground state energy}

Let us come back to the case heavy and light holes interacting via Coulomb
repulsion in the valence band of bulk III-V semiconductors.   
With the renormalized Fermi wave numbers $q_{h}$, $q_{l}$, the total kinetic
energy per particle in the ground state reads 
\begin{equation}
\frac{E_{kin}\left(q_{h},q_{l}\right)}{N}=\frac{1}{5\pi^{2}n}
\left(\frac{\hbar^{2}}{2m_{h}}q_{h}^{5}+\frac{\hbar^{2}}{2m_{l}}q_{l}^{5}\right)
\end{equation}
with the density $n$ given by Eq.~(\ref{density}), and for the total
exchange energy per particle one finds
\begin{eqnarray}
\frac{E_{ex}\left(q_{h},q_{l}\right)}{N} & = & -\frac{e^{2}}{\varepsilon_{r}}\frac{1}{16\pi^{3}n}
\Bigl(4\left(q_{h}^{4}+q_{l}^{4}\right)\nonumber\\
 & & \qquad -3\left( q_{h}^{3}-q_{l}^{3}\right)\left( q_{h}-q_{l}\right)\Bigr)\,,
\end{eqnarray}
resulting in a total energy per hole
\begin{equation}
\frac{E_{tot}\left(q_{h},q_{l}\right)}{N}=\frac{E_{kin}\left(q_{h},q_{l}\right)}{N}
+\frac{E_{ex}\left(q_{h},q_{l}\right)}{N}\,.
\end{equation}
To obtain the corresponding results for the unrenormalized Fermi wave numbers,
one just has to replace in the above expressions $q_{h/l}$ with $k_{h/l}$. Note that
the unrenormalized expression $E_{tot}(k_{h},k_{l})$ is the equivalent to first-order
perturbation theory in the Coulomb repulsion where one just computes the
expectation value of the interaction with respect to the ground state of the
non-interacting system characterized by the Fermi wave numbers $k_{h/l}$.

\begin{figure}
\centerline{\includegraphics[width=7cm]{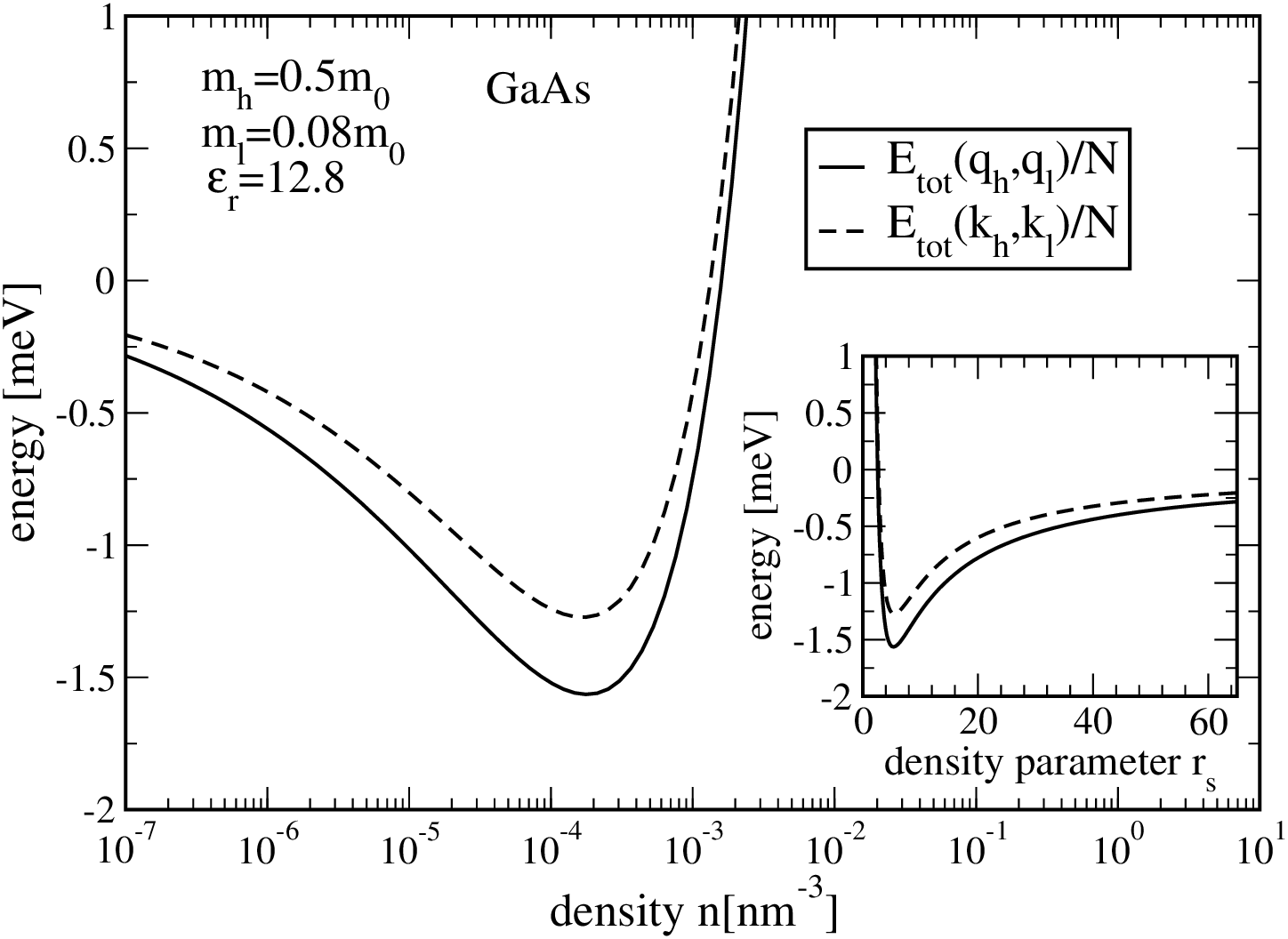}} 
\caption{ The ground state energy per particle
$E_{tot}(q_{h},q_{l})/N$ from 
the self-consistent Hartree-Fock treatment (with renormalized Fermi wave
numbers) and the result $E_{tot}(k_{h},k_{l})/N$ from 
lowest-order perturbation theory (with unrenormalized Fermi wave numbers)
as a function of the density for GaAs. The inset shows the same data as
the main panel but as a function of the density parameter $r_{s}$.
\label{fig3}}
\end{figure}

Fig.~\ref{fig3} shows the ground state energy per particle
$E_{tot}(q_{h},q_{l})/N$ from 
the self-consistent Hartree-Fock treatment and the result $E_{tot}(k_{h},k_{l})/N$ from 
lowest-order perturbation theory as a function of the density for GaAs.
As seen in the figure, it is always $E_{tot}(q_{h},q_{l})<E_{tot}(k_{h},k_{l})$, i.e. the
self-consistent Hartree-Fock approach gives the lower ground state energy.
This is clear since the ground state obtained from a self-consistent 
solution of the Hartree-Fock equations is by construction the Slater
determinant of lowest energy in the Hilbert space of the many-particle
system. Thus, any other Slater determinant state has to have a higher
energy expectation value. In fact, the renormalized Fermi wave numbers can
alternatively be obtained by minimizing $E_{tot}(q_{h},q_{l})/N$ with respect to
$q_{h}$, $q_{l}$ under the constraint of a fixed density $n$. In the absence of
interactions the minimization of $E_{kin}(q_{h},q_{l})/N$ immediately reproduces
the results of section \ref{nonint} (as it has to be), whereas for
$E_{tot}(q_{h},q_{l})/N$ one ends up with a coupled system of polynomial equations
which does not seem to allow for an explicit analytical solution.

The hole densities usually occurring in realistic samples lie at
$n=0.01\dots1.0{\rm nm}^{-3}$. At such densities, the difference between
$E_{tot}(q_{h},q_{l})/N$ and $E_{tot}(k_{h},k_{l})/N$ is indeed very small 
(cf. Fig.~\ref{fig3}),
and the density parameter $r_s$ as defined in Eq.~(\ref{denpar},(\ref{bohr})
is of order unity, giving confidence
to the validity of the Hartree-Fock approach \cite{Mahan00,Rossler04}.
At smaller densities like $n<0.001{\rm nm}^{-3}$,
$E_{tot}(q_{h},q_{l})/N$ and $E_{tot}(k_{h},k_{l})/N$ differ
appreciably. However, at these densities Hartree-Fock theory cannot be expected
to give accurate results. On the other hand, it is common in many-body 
perturbation theory
to refer to all contributions to the ground state energy beyond the
lowest-order exchange term as correlation contributions 
\cite{Mahan00,Rossler04}. In this sense the difference between
$E_{tot}(q_{h},q_{l})/N$ and $E_{tot}(k_{h},k_{l})/N$ (resulting from the 
renormalization of
Fermi momenta) can be viewed as a correlation effect.

The inset of Fig.~\ref{fig3} shows the same data as the main panel, but as 
a function of the density parameter $r_{s}$. The minimum of the
Hartree-Fock ground state energy lies at about $r_{s}\approx 5$, similarly to the case
of the usual spin-1/2 electron gas \cite{Mahan00,Rossler04}. Note that the
maximum difference between $E_{tot}(q_{h},q_{l})/N$ and $E_{tot}(k_{h},k_{l})/N$ is also
achieved around this value.

\subsection{Pair distribution function}

\begin{figure}
\centerline{\includegraphics[width=7cm]{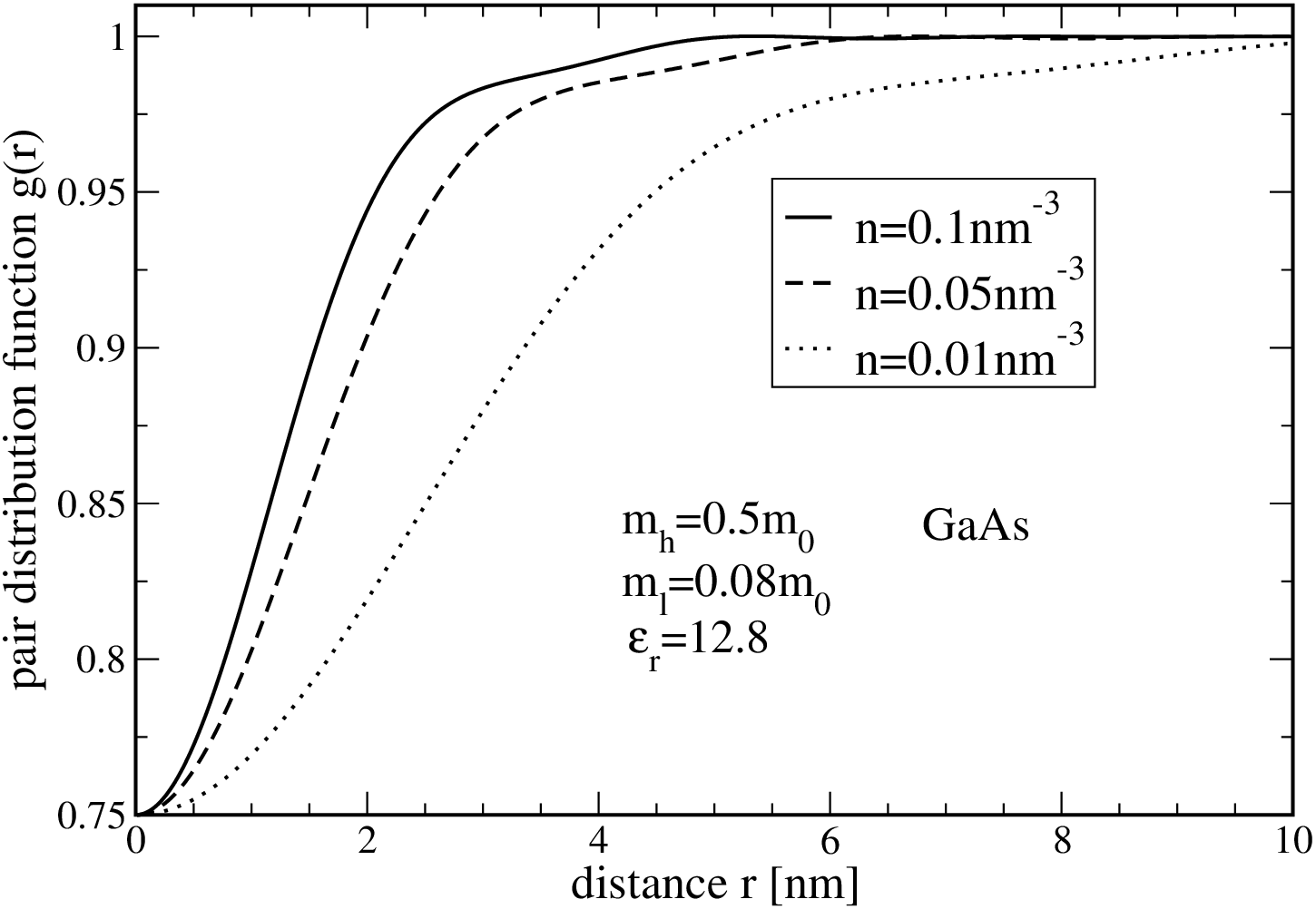}} 
\caption{The pair distribution function $g(r)$ for GaAs at three different 
densities.
\label{fig4}}
\end{figure}

It is instructive to also investigate the pair distribution function $g(r)$
defined by
\begin{equation}
n^{2}g\left(\left|\vec r-\vec r'\right|\right)=
\left\langle\sum_{I\neq J}\delta\left(\vec r-\vec r_{I}\right)\delta\left(\vec r'-\vec r_{J}\right)
\right\rangle
\end{equation}
where $I$,$J$ label the particles in the system and $\langle\cdot\rangle$ denotes the
expectation value within the ground state. The ground state obtained 
from self-consistent Hartree-Fock theory is a single Slater determinant. Here
the pair distribution function can be formulated as
\begin{equation}
g(r)=1-\left(g^{ex}_{hh}(r)+g^{ex}_{ll}(r)+g^{ex}_{hl}(r)\right)\,,
\end{equation}
where $g^{ex}_{hh}$ ($g^{ex}_{ll}$) are the exchange contributions from heavy (light)
hole states only, whereas $g^{ex}_{hl}$ stems from exchange between heavy and light 
holes. It is straightforward to calculate these contributions explicitly
using Eqs.~(\ref{overlap1})-(\ref{overlap3}). The results can be
formulated as
\begin{eqnarray}
g^{ex}_{hh}(r) & = & \left(\frac{m_h}{\tilde m}\right)^{3}
\Biggl(\frac{9}{32}\left(I_{1}(q_{h}r)\right)^{2}
+\frac{27}{16}\left(I_{2}(q_{h}r)\right)^{2}\nonumber\\
 & & \qquad+\frac{27}{8}\left(I_{3}(q_{h}r)\right)^{2}\Biggr)\,,\\
g^{ex}_{ll}(r) & = & \left(\frac{m_l}{\tilde m}\right)^{3}
\Biggl(\frac{9}{32}\left(I_{1}(q_{l}r)\right)^{2}
+\frac{27}{16}\left(I_{2}(q_{l}r)\right)^{2}\nonumber\\
 & & \qquad+\frac{27}{8}\left(I_{3}(q_{l}r)\right)^{2}\Biggr)\,,\\
g^{ex}_{ll}(r) & = & 
\left(\frac{m_h}{\tilde m}\right)^{3/2}\left(\frac{m_l}{\tilde m}\right)^{3/2}
\Biggl(\frac{27}{16}\left(I_{1}(q_{h}r)\right)\left(I_{1}(q_{l}r)\right)\nonumber\\
 & & \qquad\qquad
-\frac{27}{8}\left(I_{2}(q_{h}r)\right)\left(I_{2}(q_{l}r)\right)\nonumber\\
 & & \qquad\qquad
-\frac{27}{4}\left(I_{3}(q_{h}r)\right)\left(I_{3}(q_{l}r)\right)\Biggr)\,,
\end{eqnarray}
where we have defined
\begin{eqnarray}
I_{1}(x) & = & -\frac{\cos x}{x^{2}}+\frac{\sin x}{x^{3}}\,,\\
I_{2}(x) & = & -\frac{\sin x}{x^{3}}+\frac{1}{x^{3}}\int_{0}^{x}dy\frac{\sin y}{y}\,,\\
I_{3}(x) & = & -\frac{1}{2}\frac{\cos x}{x^{2}}+\frac{3}{2}\frac{\sin x}{x^{3}}
-\frac{1}{x^{3}}\int_{0}^{x}dy\frac{\sin y}{y}\,.
\end{eqnarray}
Note that $g(0)=3/4$, corresponding to a fermionic gas with four spin
components. In the absence of spin-orbit coupling, $q_{h}=q_{l}$,
the contributions involving $I_{2}$, $I_{3}$ cancel and one obtains
the well-known exchange terms of the usual electron gas 
\cite{Mahan00,Rossler04}.
Fig.~\ref{fig4} shows the pair distribution function in GaAs for
three different densities. At high enough hole density one can see Friedel-type
modulations of $g(r)$ whose period is essentially given
by twice the heavy-hole Fermi wave number $q_{h}$. 

\subsection{The dielectric function in random phase approximation}
\label{RPA}

Within random phase approximation (RPA), the dielectric function
is given by \cite{Mahan00,Rossler04}
\begin{equation}
\varepsilon^{RPA}(\vec k,\omega)=1-V(\vec k)\chi_{0}(\vec k,\omega)\,,
\end{equation}
where $\chi_{0}(\vec k,\omega)$ is the susceptibility of the non-interacting system.
Its real part has the form
\begin{eqnarray}
 \chi_{0}(\vec k,\omega) & = & \frac{1}{(2\pi)^{3}}\sum_{\lambda_{1},\lambda_{2}}\int d^{3}k'\Biggl[
\left|\langle\chi_{\lambda_{1}}(\vec k')|\chi_{\lambda_{2}}(\vec k'+\vec k)\rangle\right|^{2}\nonumber\\
 & & \cdot\frac{f(\vec k',\lambda_{1})-f(\vec k'+\vec k,\lambda_{2})}
{\hbar\omega-\left(\varepsilon_{\lambda_{2}}(\vec k'+\vec k)-\varepsilon_{\lambda_{1}}(\vec k')\right)}\Biggr]
\label{chi1}\\
 & = & \frac{2}{(2\pi)^{3}}\sum_{\lambda_{1},\lambda_{2}}\int d^{3}k'\Biggl[
\left|\langle\chi_{\lambda_{1}}(\vec k')|\chi_{\lambda_{2}}(\vec k'+\vec k)\rangle\right|^{2}\nonumber\\
 & & \cdot\frac{f(\vec k',\lambda_{1})
\left(\varepsilon_{\lambda_{2}}(\vec k'+\vec k)-\varepsilon_{\lambda_{1}}(\vec k')\right)}
{(\hbar\omega)^{2}-\left(\varepsilon_{\lambda_{2}}(\vec k'+\vec k)-\varepsilon_{\lambda_{1}}(\vec k')\right)^{2}}\Biggr]
\label{chi2}
\end{eqnarray}
Here $f(\vec k,\lambda_{1})$ are Fermi functions, and to obtain Eq.~(\ref{chi1}) from
Eq.~(\ref{chi2}) we have used elementary properties of the spinor overlaps
and the dispersion relations of the non-interacting system. In particular, in 
the static limit one has
\begin{eqnarray}
 \chi_{0}(\vec k,0) & = & \frac{2}{(2\pi)^{3}}\sum_{\lambda_{1},\lambda_{2}}\int d^{3}k'\Biggl[
\left|\langle\chi_{\lambda_{1}}(\vec k')|\chi_{\lambda_{2}}(\vec k'+\vec k)\rangle\right|^{2}\nonumber\\
 & & \cdot\frac{f(\vec k',\lambda_{1})}
{\varepsilon_{\lambda_{1}}(\vec k')-\varepsilon_{\lambda_{2}}(\vec k'+\vec k)}\Biggr]\,.
\label{chi3}
\end{eqnarray}
By construction, $\chi_{0}(\vec k,\omega)$ is entirely determined by the properties of 
the non-interacting system. In particular, at zero temperature, the
integration boundaries in the above expressions are given by the
unrenormalized Fermi wave numbers $k_{h}$, $k_{l}$. However, in order to be
consistent with the self-consistent Hartree-Fock treatment, one may use
the renormalized wave numbers $q_{h}$, $q_{l}$ instead. As seen above, at high 
enough densities, the difference is negligible.

For early work on dielectric response in zero-gap semiconductors
we refer to Refs~\cite{Broerman71,Broerman72}. 
An evaluation of the static expression~(\ref{chi3}) for the case of the 
two-dimensional electron gas with Rashba spin-orbit interaction has been given
by Chen and Raikh \cite{Chen99}. Their findings have recently been challenged
by Pletyukhov and Gritsev \cite{Pletyukhov05}.
The main technical obstacle there is posed
by non-elementary integrals. In the present case of the three-dimensional 
hole gas, however, the occuring integrations are mostly elementary but 
often very tedious. 

Analogously as for the three-dimensional elelctron gas 
the static dielectric function for the hole system 
at zero temperature can be formulated as
\begin{equation}
\varepsilon^{RPA}(\vec k,0)=1+\frac{k_{TF}^{2}}{k^{2}}
L\left(\frac{k}{2q_{h}},\frac{k}{2q_{l}}\right)\,,
\end{equation}
where
\begin{equation}
k_{TF}=\sqrt{\frac{6\pi e^{2}n}{\varepsilon_{r}\varepsilon_{f}}}
\end{equation}
is the usual Thomas-Fermi screening wave number.
The function $L$ is the analogue of the well-known Lindhard correction
for the electron gas. The explicit form of $L$ for the hole system, however,
is rather lengthy and tedious and shall not be given here.
For long wavelength, $k\to 0$, L approaches unity. In this
limit terms mixing heavy and light holes do not contribute to the
dielectric function. A similar finding is valid for the
plasma frequency which characterizes collective excitations at zero
wave vector \cite{Mahan00,Rossler04}. 
Here contributions mixing heavy and light holes are also
absent, and the plasma frequency $\omega_{p}$ is given by
\begin{equation}
\omega_{p}^{2}=\frac{4\pi e^{2}}{\varepsilon_{r}}\left(\frac{n_{h}}{m_{h}}+\frac{n_{l}}{m_{l}}\right)\,,
\end{equation}
where $n_{h}$, $n_{l}$ are the densities of heavy and light holes, respectively.

\section{Discussion and outlook}
\label{discussion} 

We have studied the homogeneous interacting hole gas in 
$p$-doped bulk III-V semiconductors modelled by Luttinger's Hamiltonian
in the spherical approximation. The Coulomb repulsion is taken into account 
via a self-consistent
Hartree-Fock treatment. As a nontrivial feature of the model, the 
self-consistent solutions of the Hartree-Fock equations 
can be found in an almost
purely analytical fashion. As we have discussed in detail in section
\ref{comparison} this is not the case for other types of effective
spin-orbit coupling terms. 
As an important qualitative feature, the Coulomb repulsion 
renormalizes the Fermi wave numbers for heavy and light holes: The 
interaction leads to a redistribution of occupation numbers from
heavy holes to light holes compared to the non-interacting case.
As a consequence, the ground state energy found in the self-consistent 
Hartree-Fock approach and the result from lowest-order perturbation theory 
differ from each other. By construction, the self-consistent Hartree-Fock
result gives the lower ground state energy.

The three-dimensional III-V semiconductor hole gas in particularly relevant
for ferromagnetic semiconductors which are usually $p$-doped materials. 
In the theoretical description of
these materials, the interaction between itinerant charge carriers is
most often neglected \cite{Timm03,MacDonald05,Jungwirth06}, or absorbed in
effective Fermi liquid parameters \cite{Dietl01}. The typical hole densities
in Mn-doped GaAs, the most prominent and best-studied ferromagnetic 
semiconductor, are of order $n\approx 0.1{\rm nm}^{-3}$. For such carrier 
concentrations, the density parameter $r_s$ 
is of order unity, giving confidence
to the validity of the Hartree-Fock approach \cite{Mahan00,Rossler04}.
Moreover, as seen in the present investigation, the renormalization of 
Fermi wave numbers is negligible at such densities, i.e. the 
interacting ground state in Hartree-Fock approximation and the non-interacting
ground state are practically the same. In this sense, the abovementioned models
for ferromagnetic semiconductors neglecting the Coulomb interaction are 
supported by the present study. However, the single-particle Hamiltonian
used there is a simplified one which does not take into account the
split-off band. A more complete description of the valence band including 
these states is given by the six-band Kohn-Luttinger model \cite{Luttinger55}
as used in Refs.~\cite{Dietl00,Dietl01}.
In fact, the influence of the split-off band is known to be important
for the stability of the ferromagnetic order \cite{Konig01}.
However, for the full six-band Kohn-Luttinger Hamiltonian mainly analytical
progress like in the present work is certainly not possible, and one 
would need  to resort to more complicated numerics

\begin{figure}
\centerline{\includegraphics[width=7cm]{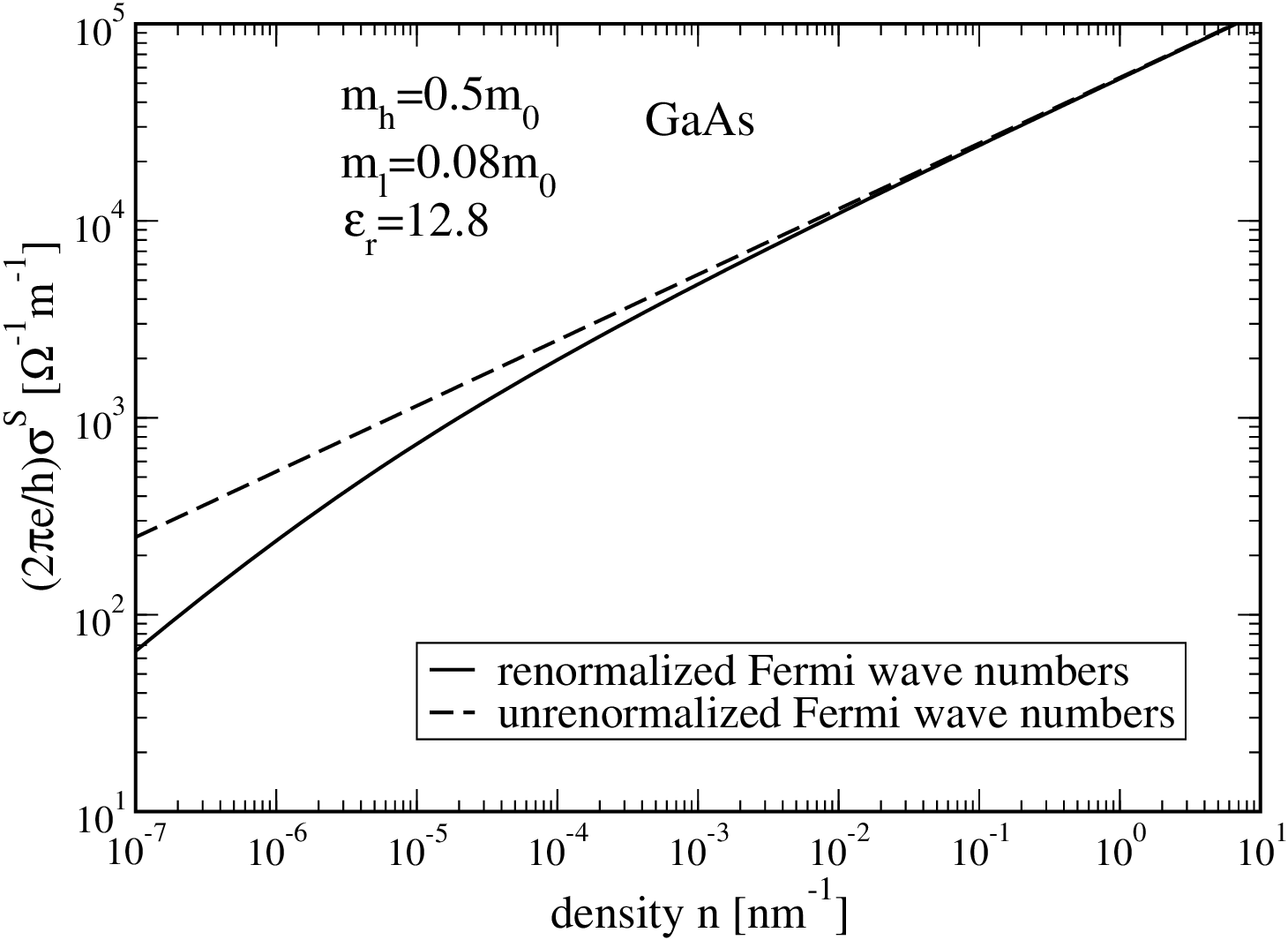}} 
\caption{The spin-Hall conductivity $\sigma^{S}$ (converted to units of charge 
transport) as a function of hole density $n$ for GaAs.
\label{fig5}}
\end{figure}

During the last years various predictions and subsequent experiments
regarding spin-Hall transport
in semiconductor systems have attracted a very remarkable deal of interest
\cite{Murakami03,Sinova03,Schliemann05}; for a recent overview see also
Ref.~\cite{Schliemann06}. The first work opening the field of intrinsic
spin-Hall effect was a paper by Murakami, Nagaosa, and Zhang who considered
a $p$-type bulk III-V semiconductor \cite{Murakami03}. The single-particle
Hamiltonian used in Ref.~\cite{Murakami03} is the same as here with the
Coulomb repulsion between the holes being neglected \cite{Murakami03}.
For a disorder-free system, the spin-Hall conductivity is given by
\cite{Schliemann06,Schliemann04}
\begin{equation}
\sigma^{S}\left(q_{h},q_{l}\right)=\frac{e}{4\pi^{2}}\frac{\gamma_{1}+2\gamma_{2}}{\gamma_{2}}
\left(q_{h}-q_{l}\right)\,,
\end{equation}
where the direction of the spin current, its polarization direction, and 
the direction of the electric field are mutually perpendicular.
In the above expression we have used the renormalized Fermi wave numbers
$q_{h}$, $q_{l}$. Figure \ref{fig5} shows a spin-Hall conductivity as a function
of hole density for GaAs both for renormalized and unrenormalized
Fermi wave numbers. To facilitate the comparison to the usual charge
conductivity, we have converted the spin-Hall conductivity to units
of charge transport by multiplying with a factor of $e/\hbar$. As shown in the 
figure, at densities $n\gtrsim 0.01{\rm nm}^{-3}$ typical for realistic
samples, the difference between the case of renormalized and unrenormalized 
wave numbers is negligible. Appreciable discrepancies occur only at small 
densities, where the validity of the Hartree-Fock treatment becomes 
questionable anyway.

Finally, we hope that the present study will initiate further investigations
on interacting III-V semiconductor hole systems. One possible
direction is to perform (presumably numerical) Hartree-Fock calculations
for more complex band structure models as mentioned above. Another obvious
goal for future studies is to investigate many-body effects beyond the
Hartree-Fock level.

\acknowledgments{This work was supported by the SFB 689 
``Spin Phenomena in reduced Dimensions''.}

\begin{table}
\begin{tabular}{c|c|c|c|c|c|}
  & $\gamma_{1}$ & $\gamma_{2}$ & $\frac{m_{h}}{m_{0}}$ &  $\frac{m_{l}}{m_{0}}$ & $\varepsilon_{r}$\\ 
\hline
GaAs & 7 & 2.5 & 0.5 & 0.08 & 12.8 \\
InAs & 20 & 9 & 0.5 & 0.026 & 14.5 \\
InSb & 35 & 15   & 0.2 & 0.015 & 18.0 \\
\end{tabular}
\caption{The Luttinger parameters $\gamma_{1}$, $\gamma_{2}$, effective hole masses
$m_{h/l}$, and static dielectric constants $\varepsilon_{r}$ for the III-V semiconductors
GaAs, InAs, and InSb.
\label{table1}}
\end{table}

\end{document}